# Quasi-HfO$_x$/ AlO$_y$ and AlO$_y$/ HfO$_x$ Based Memristor Devices: Role of Bi-layered Oxides in Digital Set and Analog Reset Switching

*Pradip Basnet*[†]*, *Erik Anderson*[‡], *Bhaswar Chakrabarti*[¶], *Matthew P. West*[†], *Fabia Farlin Athena*[†] *and Eric M. Vogel*[†]

[†]School of Materials Science and Engineering, Georgica Institute of Technology, Atlanta, Georgia 30332, USA
[‡]George W. Woodruff School of Mechanical Engineering, Georgia Institute of Technology, Atlanta, GA, 30313, USA
[¶]Department of Electrical Engineering, Indian Institute of Technology Madras, Chennai, Tamil Nadu- 600036, India
*Address correspondence to: pbasnet6@gatech.edu

**ABSTRACT**

*Understanding the resistive switching behavior, or the resistance change, of oxide based memristor devices is critical to predict their responses with known electrical inputs. Also, with the known electrical response of a memristor, one can confirm its usefulness in non-volatile memory and/or in artificial neural networks. Although bi- or multi-layered oxides have been reported to improve the switching performance, compared to the single oxide layer, the detailed explanation about why the switching can easily be improved for some oxides combinations is still missing. Herein, we fabricated two types of bi-layered heterostructure devices, quasi-HfOx/AlOy and AlOy/HfOx sandwiched between Au electrodes, and their electrical responses are investigated. For deeper understanding of the switching mechanism, performance of a HfOx only device is also considered, which serves as a control device. The role of bi-layered heterostructures is investigated using both the experimental and simulated results. Our results suggest that synergistic switching performance can be achieved with a proper combination of these materials and/or devices. These results open the avenue for designing more efficient double- or multi-layers memristor devices for analog response.*

**INTRODUCTION**

Due to the increasing use of data in the recent era, memory and device improvement became the highest priority to continue using the current digital computing or von Neumann architecture. At the same time, alternative technologies for future non-volatile memory (NVM) are attracting a great deal of attention as the scalability of the conventional flash memory is approaching its physical limits.[1-3] As such, the resistive random-access memory (ReRAM) devices have emerged as a promising candidate in the field of both NVM and neuromorphic computing applications.[2, 4-7] ReRAM has various properties outweighing the transistor-type memories integrated to the digital computing, for example, high-speed operation (sub-ns), low power consumption (<0.1 pJ), high endurance (> 10$^{10}$ cycles), and high cell density.[1-2, 8-9]

An oxide-based ReRAM device consists of a capacitor-like metal-insulator-metal (M-I-M) structure. It is widely accepted that the switching performance of a memristor—also called a resistance switching device—is based on electron-ion dynamics.[3-4, 10-11] The resistance change between the high resistance state (HRS) and the low resistance state (LRS) is attributed to formation and rupture of the conducting filament(s) (CFs) in the insulator layer, also known as active layer.[3, 11-12] This change of conductance, *via* migration of charge species-- including oxygen ions or vacancies, bears similarity with the adaptation of synaptic weights that enables the biological systems to learn and function.[5, 13] In the recent years, various oxide based memristor devices have been proposed to emulate the brain in analogous to the biological synapse, which is similar to the two-terminal M-I-M structure of ReRAM devices.[3, 14-15] Transition metal oxides such as HfO$_2$, TiO$_2$, Ta$_2$O$_5$, Nb$_2$O$_5$ and ZrO$_2$ are some of the most studied materials for NVM and brain-inspired computing applications.[3, 6, 12, 16] In the literature, materials modifications by both doping (e.g. Si-doped Ta$_2$O$_5$, Al-doped HfO$_2$)[3, 17] and stacking the oxides layers (e.g. Al$_2$O$_3$/TiO$_2$; HfO$_x$/HfO$_2$; HfO$_x$/TiO$_2$; Al$_2$O$_3$/Nb$_2$O$_5$; and HfO$_x$/AlO$_y$)[4, 6-7, 14-15] have been suggested to improve the switching performance characteristics, such as low operating voltage, wide memory window, and high endurance cycles. Recently, Tan et. *al*. and Kim et. *al*. separately reported the multilevel conductance behavior of HfO$_x$/HfO$_2$ homo-bilayer structure and temperature dependent spike-induced synaptic behavior of HfO$_x$/AlO$_y$ respectively.[14-15] Although improved resistive switching performance has been reported for bi- or multi-layered structure devices compared to single-layer devices, the true origin and underlying physical mechanism is still unclear. It is recognized that not only the electrodes and active layer(s) make an impact on device switching performance,[7, 18] but also the oxide interface(s) play a significant role.[4, 8, 15, 17] Both filamentary-dominated (digital I-V) and interface-dominated (analog I-V) switching characteristics have been reported for bi- and multi-layered $HfO_x$ based devices.[14-15] Nevertheless, there is very limited information about the material-dependent conduction mechanisms of each layer within a bi- or multi-layered oxides devices and the applied electrical inputs.

In this work, we report the switching performance of quasi- $HfO_x/AlO_y$ and $AlO_y/HfO_x$ memristor devices, fabricated on thermally oxidized $SiO_2$ (280 nm)/Si wafers. We further analyze the differences in the experimental I-V curves that are based on the structure and properties of both tunneling ($AlO_y$) and the active layer ($HfO_x$) — for example, electron affinity ($\chi$) and interfacial distance. More specifically, simulations using known thicknesses of metals and oxides, literature-reported energy band levels of oxides, and the oxide-metal junctions, result in different tunnel current densities between different device structures. The electron tunneling is estimated for the double barrier insulators (M-I-I-M) structure, as described in the literature.[19-21] We then explore the possible conduction mechanisms based on both energy band levels and the experimental I-V curves. We report the optimized post-annealing conditions for each device and compare the electrical performances for digital set and analog reset under the optimized electrical inputs. Finally, we discuss the useful applications of these devices for both NVM and synapses in the artificial neural networks.

**EXPERIMENTAL SECTION**

**Device fabrication and characterization.** We fabricated three devices We fabricated three devices that have different oxide layers: (1) $HfO_x$(2.5 nm)/$AlO_y$(2.5 nm), (2) $AlO_y$(2.5 nm)/$HfO_x$(2.5 nm), and (3) $HfO_x$(5 nm). Each device has a Ti capping layer and Au top- and bottom-electrodes. The quasi- $HfO_x/AlO_y$ and $AlO_y/HfO_x$ device structures are illustrated in Fig. 1. Hereafter we denote these devices as D1 and D2, respectively. The $HfO_x$-only structure is used as a reference, (Ref) device. All three devices were fabricated using the same Atomic Layer Deposition (ALD) system and lift-off photolithography. For further experimental details about making the devices, we refer to our recently published article.[11] Briefly, 250 °C thermal deposition was used to deposit both $HfO_x$ and $AlO_y$ layers using Tetrakis (dimethylamido) hafnium (TDMA-Hf) and trimethylaluminum (TMA) precursors, respectively. The de-ionized (DI) water was used as the oxygen precursor in both cases. The pulse times for all deposition cycles were fixed at 0.06 s for both DI water and TMA and 0.25 s for TDMA-Hf. Also, the number of deposition cycles were fixed to 22 for both layers, and it was based on the optimized deposition rates to be 2.5 ± 0.2 nm (with uniformity 2-3%). To avoid the possible undesirable variations, all devices were fabricated at the same time on the same size $SiO_2$(280 nm)/Si (= 1"x 1" square) substrates from the same wafer. As-prepared materials were characterized for chemical compositions using X-ray photoelectron spectroscopy (XPS). A Keithley 4200-SCS system was used for electrical properties measurements at room temperature.[11] All devices were characterized by probing the top electrode (TE) while the bottom electrode (BE) was grounded, shown in Fig. 1 inset. Also, to confirm the effect of heat at $HO_x-AlO_y$ interface or device performances, post-annealing treatment was performed to both the devices D1 and D2 by annealing at 400 °C and 500 °C for 5 mins, in the fixed $N_2$ flow (=10 sccm) using SSI Rapid Thermal Processor system.

**RESULTS AND DISCUSSION**

Fig. 1 shows the experimental I-V curves, of D1-annealed at 400 °C, and as prepared D2, devices recorded after the CFs stabilization. For the detailed stabilization process of these devices, we refer to the Fig. S1 in the Supplementary Information (SI). Note that all the results we present here are for annealed D1 and as-prepared D2 devices, unless otherwise stated; these represent the best performance devices prepared in the above-mentioned experimental conditions. It is well established that the forming and/or pre-stabilization steps are mandatory for the oxide based filamentary memristors depending on the oxide thickness. In our case, the thickness of $HfO_x$ ~ 3 nm or above, and other parameters as described in the literature.[11, 22-24] It is noteworthy that the D1 devices do not require the filament formation but only the stabilization to get reasonably stable I-V characteristic curves. The plausible pre-existing conductive path(s), that is (are) formed due to annealing (at 400 °C), can be regarded as the main source resulting these devices to be forming-free. In contrast, the D2 devices require a low forming voltage, Vf ~ 3 V, and stabilization (*see* Fig. S1 and S2 in the SI). Also, the use of Icc (current compliance) especially on electroforming and set is critical: without the Icc the I-V characteristic curves becomes stochastic, leading to permanent breakdown of the devices. This could be attributed to the fact that the formed CFs might be easily and permanently damaged with higher electrical currents. Furthermore, it is interesting that the devices D1 (D2) set at negative (positive) bias relatively at lower Icc (= 0.1 mA). We propose that this dissimilar electrical response is based on not only the differences of oxygen vacancies in the oxide layers but also the asymmetry in tunnel current density coming from the different barrier heights of $HfO_x$ and $AlO_y$. The role of barrier heights will be discussed later in detail. The role of oxygen vacancies is examined using the XPS results, which showed that the $HfO_x$ is relatively more sub-stoichiometric (than the $AlO_y$) as O:Hf and O;Al ratios were estimated to be 1.8 and 1.4, respectively.

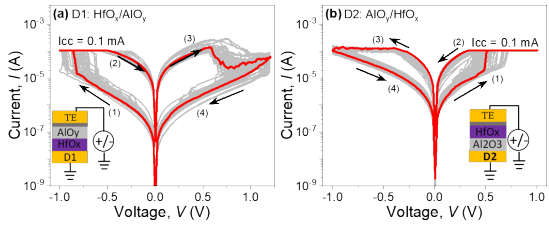

Fig. 1. (Color online) Typical resistive switching, or non-linear I-V characteristics, of quasi-$HfO_x/AlO_y$ (D1- annealed at 400 °C) and $AlO_y/HfO_x$ (D2- as prepared) devices under the same positive and negative biasing conditions. Both test runs were 100+ cycles: (1)→(2)→(3)→(4). Results show that both the devices exhibit digital set and analog reset, steps (1) and (3) respectively.

This means that the oxygen vacancy concentration in $HfO_x$ is higher (10%) than in $AlO_y$ (~6.7%). These values suggest that the device materials were sub-stoichiometric $HfO_{1.8}$ and $Al_2O_{2.8}$. In this scenario, at higher negative voltage the positively charged oxygen

D1 since the $HfO_x$ layer is less stoichiometric (than the $AlO_y$ layer on top). This also explains why these devices turn on with the applied negative bias. As shown, the device D1 showed a reasonably stable switching for the tested 100+ set-reset cycles, recorded in a continue "auto mode" data acquisition setting. Based on these results, it is evident that the $AlO_y$ layer is playing a vital role in switching: the forming-free and low power switching can be attributed to the series combination of $HfO_x/AlO_y$ interface, which has an oxygen vacancy gradient, and the $AlO_y$ layer serves as a tunneling resistor in series.[3, 25] On the other hand, D2 devices set with positive voltage that instead pulls oxygen ions towards the $HfO_x$/Ti interface after forming the CF(s) in the active $HfO_x$ layer. This switching difference should also be based on how the CF(s) is(are) formed in the active layer: For example, conduction of the filament(s) in the D1 devices can be controlled by both oxygen vacancies present and Al diffusion into the $HfO_x$ layer, while only the oxygen vacancies are expected in the filament(s) of D2 devices as they were not annealed. Since the device D2 has $HfO_x$ layer on top, the CF(s) formation should start from the TE with positive bias. Intrinsically, the $AlO_y$ layer is more resistive than a comparably thick $HfO_x$ layer, predominantly due to the tunneling resistance that arises from the higher potential barrier of $AlO_y$. Therefore, a greater voltage drop will manifest in the $AlO_y$ layer with the remaining voltage occurring in the active or $HfO_x$ layer. In this case, the role of the $AlO_y$ is only to reduce the power of the devices by increasing the resistance.

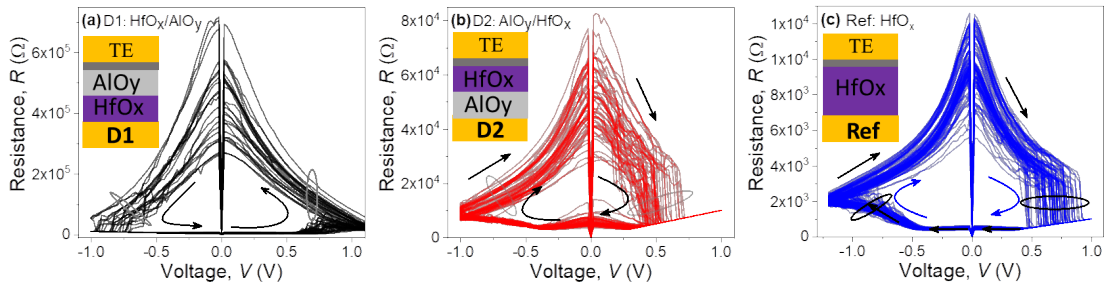

Fig. 2. (Color online) Resistive switching performance, or resistance-voltage plots, of the quasi-$HfO_x/AlO_y$ (D1- annealed at 400 °C) and $AlO_y/HfO_x$ (D2- as prepared) devices (figs. a & b), compared with a $HfO_2$-only reference (Ref) device (fig. c). Note that the D1 device exhibit remarkable improvement in switching -including more gradual resistance change and the low power operation.

vacancies from the $HfO_x$ layer, in the D1 devices, are pulled towards the $AlO_y$ layer or $AlO_y$/Ti interface, creating metal-rich and electronically conductive channel(s). In other words, the stable CFs formation starts from the bottom electrode in the case of device

The resistive switching behaviors of D1 and D2 devices are compared with the Ref device in Fig. 2. Results suggest that the D1 exhibits more gradual resistance change both in set and reset when compared to the D2 and Ref devices. This result agrees well with variations than the LRS values, *see* Fig. 3(a). This could be due to the random change in the size, and perhaps the geometry, of the CFs during re-oxidation process, which we have also observed with the Ref or amorphous 5 nm $HfO_x$ devices. As seen from Fig. 3(b), the HRS/LRS ratio of the D2 devices decreased by almost one order of magnitude. We suggest that this is due to the higher leakage current at HRS, *see* Fig. 1(b). Nevertheless, the memory window of D2 (i.e. HRS to LRS ratio ~10) is still large enough for the ReRAM application as reported in the literature.[26, 28-30] Note that for D2, this ratio is comparable with the value of the $HfO_2$ reference device (data not shown here). Hence, the observed forming-free, switching at low set/reset the literature, reported for $HfO_x$-Al matrix.[26-27] Thus, based on the results, it can be inferred that annealing the D1 devices at 400 °C is beneficial in helping diffuse Al into the $HfO_x$ layer. Consequently, annealing stabilizes the generation of CFs and helps improve the resistive switching uniformity by reducing the oxygen vacancy formation energy. Further, we observed an even wider memory window with the D1 devices when the Icc is increased to 1.0 mA as shown in Figs. S3 – S4 in the SI.

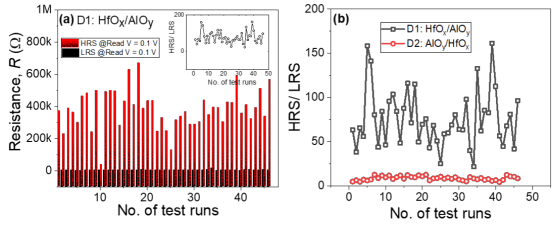

*Fig. 3. (Color online) (a) The HRS (high resistance state) and LRS (low resistance state) values comparisons of D1 and (b) HRS/LRS ratios comparisons of the D1 and D2 devices. It is observed that the HRS/LRS ratio of D1 is significantly higher (~50 to 100) as compared to D2 (~10).*

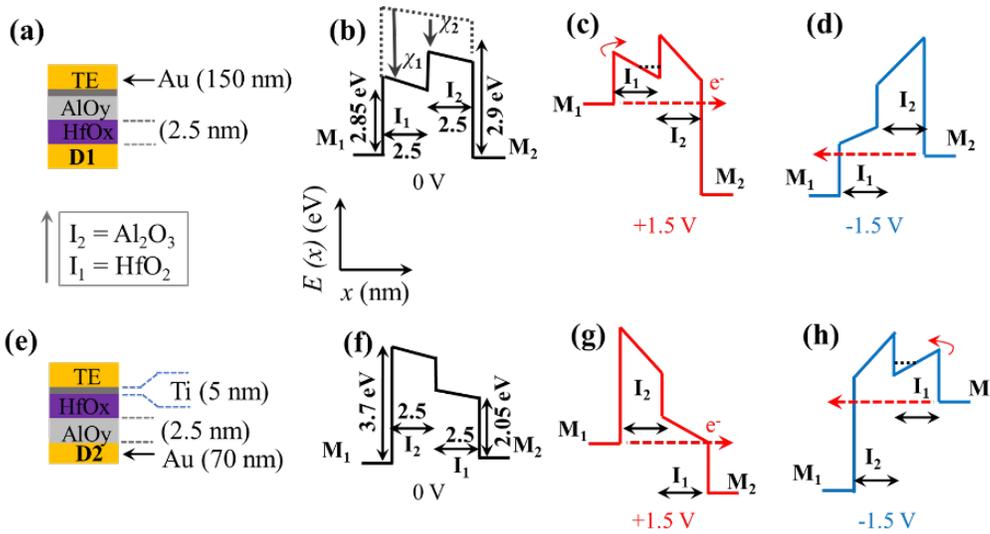

*Fig. 4. (Color online) Device structures and the simulated band diagrams of D1 (top-row) and D2 (bottom row). Note that the actual thicknesses of both insulator layers $HfO_2$ and $Al_2O_3$ were used for calculation. The top electrode (TE) was biased, and the bottom electrode (BE) was grounded. Results shows that step tunneling and resonant tunneling can occur in either device structure depending on bias polarity and magnitude. As labeled, (b) and (f) show equilibrium at zero bias, (c) and (g) at forward bias, and (d) and (h) at reverse bias.*

Fig. 3 show the comparisons of resistance values in HRS and LRS of D1, along with the ratios of HRS to LRS of both D1 and D2. Statistically, it is observed that the HRS values exhibit relatively more "cycle-to-cycle" voltage with reasonable switching cycles, and great a memory window of D1 is a remarkable improvement in the switching performance.

To further appreciate the observed differences in switching behavior of D1 and D2, we simulate the electron tunneling current densities, $J(V)$, in these M-I-I-M structures using the transfer matrix method (TMM).[31] The purpose of the simulations is to highlight how the differences in the M-I-M and M-I-I-M structures used here can potentially affect the electrical response. For tunneling of electrons through a potential barrier, current density is a combination of tunneling probability and the occupational density of the electron energy levels, given by the following expression:

$$J(V) = J_{(BE \to TE)} - J_{(TE \to BE)} =$$
$$\frac{4\pi m_e^* e}{h^3} \int_0^\infty T(E_x) \left[ \int_{E_x}^\infty f_L(E) - f_R(E+eV) dE \right] dE_x \text{ -- (1)}$$

where $e$, $\hbar$, $m_e^*$ are the electronic charge, Planck's constant, and electron effective mass, respec. $E_x$ and $E$ are longitudinal and total electron energy, respectively. $f(E) = \frac{1}{1+\exp\left(\frac{E-E_f}{k_b T}\right)}$ is the Fermi-Dirac distribution with $E_f$ as electrode Fermi energy level, $k_b$ is the Boltzmann constant, and $T$ is the absolute temperature (300 K for calculation). The term $f_{BE}(E) - f_{TE}(E+eV)$ marks the difference in occupational probabilities between an electron travelling from one electrode to other reversibly. The tunneling current is simulated by following the approach outlined in Ref. [31]. The transmission probability, $T(E_x)$ is numerically calculated by segmenting the M-I-I-M potential barrier into discrete rectangular slices. We solve Schrödinger's equation for each constant potential region then apply continuity boundary conditions between the interfaces. This produces a matrix that represents the transmission between each barrier slice, giving the overall transmission through the full potential barrier $T(E_x)$. In generating the M-I-I-M tunneling barriers the following reported properties were used: electron affinity, $\chi$ = 2.25 eV, band gap energy, $E_g$ = 5.6 eV, and dielectric constant, $k$ = 20 for $HfO_2$ and $\chi$ = 1.4 eV, $E_g$ = 6.4 eV, and $k$ = 7.6 for $Al_2O_3$.[19] Likewise, well-known values of work function, $\Phi$ = 4.3 eV for Ti and $\Phi$ = 5.1 eV for Au were used. Fig. 4 show illustrated band diagrams for D1 and D2 alongside schematics of the device structures. These barrier structures suggest that high tunneling nonlinearity and asymmetry may occur through step tunnelling mechanisms.[20,32]

Fig 5 shows the simulated tunneling current density $J(V)$ for the M-I-I-M devices. Both D1 and D2 devices exhibit highly amplified asymmetric $J(V)$ characteristics relative to the single barrier reference device. This can be linked to step tunneling, qualitatively depicted in Fig. 4 (d, g). In D1, for instance, at sufficiently large negative bias the conduction band profile is bent to where electrons tunnel through an effectively shortened distance through only the $Al_2O_3$ layer. The onset of this step tunneling in D1 is observed at -2.3 V, as depicted by the rapid increase in reverse bias current and nonlinearity. In contrast, the step tunneling of D2 manifests at +2.0 V. The flip in the polarity in which asymmetry occurs in this device is a clear indicator of the reversed ordering of the $HfO_2$ and $Al_2O_3$ layers. The earlier onset is likely due to the 0.8 eV Au/Ti electrode work function contrast, therein lowering the voltage needed for step tunneling to occur.[33] We also point out an absence of resonant tunneling in these structures as the potential barriers were too tall to permit a resonant well to form at low enough energy to contribute resonant behavior.

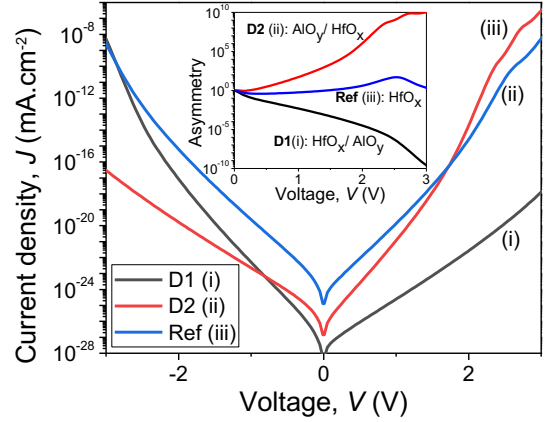

Fig. 5. (Color online) Simulated J-V characteristics of D1 and D2 against a reference device at room temperature (T = 300 K). (Inset) Asymmetry (= $\left|\frac{I(+V)}{I(-V)}\right|$) as a function of bias. The device responses are significantly controlled by the energy band positions of the insulators in the M-I-I-M structure, resulting in non-linear and asymmetric behavior.

The overall implications of these simulations indicate that under strictly tunneling-based electron conduction, the I-V curves of D1 and D2 are expected to drastically differ. It is noteworthy that the presence or absence of defects in the active and/or barrier layer also plays a key role to change the current through the devices. We posit that the enhanced tunneling currents impact the forming mechanism as well as set switching. The estimated higher tunneling currents of the D1 and D2 devices under negative and positive bias respectively agree well with the observed set switching polarity of these devices. The overall switching mechanism of these devices, however, can be complex once the CF(s) are formed, or at least we will need to include both filament/ ionic contribution in combination with tunneling simulation. It is worth mentioning during HRS state the electrical response should be predominantly governed by tunneling of the electrons, as modeled by TMM. As the tunneling simulations only take into account electron tunneling through ideal insulator regions, the predicted *J(V)* curves show a drastic range of current density magnitudes. However, practical devices are more likely to experience alternative electron conduction mechanisms as well, including Poole-Frankel emission or defect-dominated tunneling mechanisms as well as natural spatial variation in the dielectric layer. Thus, the higher experimental current of these devices could be due to the oxygen vacancy gradient present through the HfOx/AlOy interface (*see* Fig. S5 in the SI), which can cause a lowering of the effective tunneling barrier thickness in either layer.

**CONCLUSION**

In summary, we report a simple and scalable method to fabricate a bi-layered $HfO_x$-based crossbar memristor. Results from the electrical measurements and simulation showed that the position of active or barrier layer in the bi-layered structured device is critical. For example, inserting a sub-stoichiometric 2.5 nm $AlO_y$ barrier layer between the 2.5 nm active layer $HfO_x$ and the Ti capping electrode layer showed the best switching performance. The performance degraded noticeably when swapping the active and barrier layers. Including a control device of 5 nm $HfO_x$, all three devices tested here showed the digital set and analog reset switching. We propose that the switching performance of these bi-layered memristor devices is controlled by the combination of electron- and ion-based conductions. Based on the observed analog switching with both positive and negative biasing and improved switching performance, we infer that these devices can be good candidates as synapses for neuromorphic computing. In addition to the requirement of device size and behavior, however, it may require an extensive study to further improve the analog response of these devices for commercial production.

**Acknowledgement**. This work is supported by the Air Force Office of Scientific Research MURI entitled, "Cross-disciplinary Electronic-ionic Research Enabling Biologically Realistic Autonomous Learning (CEREBRAL)" under award number FA9550-18-1-0024. The device materials depositions as well as the device fabrication were performed in the Cleanroom at the Georgia Tech Institute for Electronics and Nanotechnology (IEN), Atlanta, GA, USA, which is a member of the National Nanotechnology Coordinated Infrastructure and is supported by the National Science Foundation (grant ECCS1542174).

*Notes*: The authors declare no competing financial interest.

***See the Supplementary Information (SI) for***: Forming and pre-stabilization of quasi- $HfO_x$/$AlO_y$ and $AlO_y$/$HfO_x$ devices; Switching performances of quasi- $HfO_x$/$AlO_y$ and $AlO_y$/$HfO_x$ devices with the possible conduction mechanisms at HRS and LRS; and XPS depth profile of the device materials: $HfO_x$/$AlO_y$ (annealed at 400 °C) and $AlO_y$/$HfO_x$ (as-prepared) thin films. This material is available free of charge *via* the Internet at https://pubs.acs.org/